\documentclass{article}

\usepackage{amsmath, amssymb, textcomp}
\usepackage[a4paper,twoside,inner=75pt,outer=75pt]{geometry} 
\usepackage[english]{babel}
\usepackage{parskip}
\usepackage{color}
\usepackage{enumerate}
\usepackage{graphicx}
\usepackage{layout}
\usepackage[sort&compress,numbers,square]{natbib}
\usepackage{fancyhdr}
\usepackage{caption}
\usepackage{upgreek}
\usepackage{multirow}
\usepackage{gensymb}
\usepackage{titling}
\usepackage{multicol}
\usepackage{wrapfig}
\newcommand{\bpt}{\color{black}}
\usepackage{soul}

\title{Universal scaling of flow curves: comparison between experiments and simulations}
\date{\today}
\author{Riande I. Dekker\thanks{Corresponding author: r.i.dekker@uva.nl} \thanks{Van der Waals-Zeeman Institute, Institute of Physics, University of Amsterdam, Science Park 904, 1098 XH Amsterdam, The Netherlands},  Maureen Dinkgreve\footnote[1]{}, Henri de Cagny\footnote[1]{} \thanks{Unilever Research Laboratories, Vlaardingen, The Netherlands}, Dion Koeze\thanks{Process \& Energy Laboratory, Delft University of Technology, Leeghwaterstraat 39, 2628 CB Delft, The Netherlands},\\ Brian P. Tighe\footnote[4]\\  and Daniel Bonn\footnote[1]{}}

\begin{document}

\maketitle

\begin{abstract}
Yield stress materials form an interesting class of materials that behave like solids at small stresses, but start to flow once a critical stress is exceeded. It has already been reported both in experimental and simulation work that flow curves of different yield stress materials can be scaled with the distance to jamming or with the confining pressure. However, different scaling exponents are found between experiments and simulations. In this paper we identify sources of this discrepancy. We numerically relate the volume fraction with the confining pressure and discuss the similarities and differences between rotational and oscillatory measurements. Whereas simulations are performed in the elastic response regime close to the jamming transition and with very small amplitudes to calculate the scaling exponents, these conditions are hardly possible to achieve experimentally. Measurements are often performed far away from the critical volume fraction and at large amplitudes. We show that these differences are the underlying reason for the different exponents for rescaling flow curves. 

\end{abstract}

\begin{multicols}{2}

\section{Introduction}
Complex fluids, such as emulsions, suspensions, foams and pastes, form an important class of materials, exhibiting both solid- and liquid-like behaviour. Understanding and predicting the flow behaviour of these complex materials is of industrial and fundamental importance \cite{Larson1999, Denn2011}. These materials show the emergence of a yield stress for volume fractions above a critical volume fraction $\phi_{c}$, called the jamming transition. At small stresses yield stress materials behave like solids, deforming in an elastic manner. However, once a critical stress, called the yield stress ($\sigma_{y}$), is exceeded, the material starts to flow. Describing the flow properties of yield stress materials as a function of the volume fraction has become an important research topic, both in experimental \cite{Nordstrom2010,Paredes2013,Dinkgreve2015} and simulation work \cite{OHern2003,Vagberg2014,Dagois-Bohy2017}. Universal rescaling of yield stress flow curves would enable us to predict the yield stress of a material when the volume fraction and surface tension are known \cite{Paredes2013}. 
Flow curves for concentrated systems with $\phi > \phi_{c}$ can be described by the Herschel-Bulkley model \cite{Herschel1926}
\begin{equation}
\sigma = \sigma_{y} + K \dot{\gamma}^\beta
\end{equation}  
where $\sigma$ is the stress, $\dot{\gamma}$ is the shear rate and $K$ and $\beta$ are model parameters. The vanishing of the yield stress with decreasing volume fraction can be described as a power law in the distance to jamming
\begin{equation}
\sigma_{y}=\sigma_{0}\lvert \Delta \phi \rvert^{\Delta}
\end{equation} 
with $\sigma_{0}$ {\bpt a constant on the order of} the Laplace pressure of the droplets and $\Delta \phi = \phi-\phi_{c}$. 

Paredes et al. \cite{Paredes2013} investigated the flow properties of one such yield stress fluid: an emulsion with volume fractions above and below the jamming transition. They were able to scale all flow curves with respect to the volume fraction into two master curves, one for the supercritical and one for the subcritical volume fractions, by plotting $\sigma/\lvert \Delta\phi \rvert^\Delta$ versus $\dot{\gamma}/\lvert \Delta\phi \rvert^\Gamma$. Similar scaling values were found by Nordstrom et al. \cite{Nordstrom2010} and Basu et al. \cite{Basu2014} when scaling flow curves of a soft-colloid system.
Paredes et al. interpreted scaling collapse as evidence for a critical transition in the dynamics, from liquid- to solid-like behaviour.  
In the analogy with equilibrium phase transitions, they supposed that the scaling exponent values are universal, i.e. independent of particle interactions. 
{\bpt Subsequent experimental studies provided support for this hypothesis:} Dinkgreve et al. \cite{Dinkgreve2015} found that experimental flow properties of other soft sphere systems, with different interparticle interactions, could also be scaled with power laws in the distance to jamming. The scaling parameters for all different systems have, within numerical uncertainty, the same values of $\Delta \approx 2$ and $\Gamma \approx 4$. 

{\bpt Simulations of viscous soft spheres are able to approach much closer to the jamming point than experiments: they can easily reach excess volume fractions, strain amplitudes, and dimensionless strain rates on the order of $10^{-6}$ or lower, which is orders of magnitude smaller than typical experimental lower bounds. As a result, simulations and experiments have typically probed different windows of response. And indeed, numerical estimates of critical exponents have tended to differ from experiments. For example, the yield stress exponent $\Delta$ is generally estimated to be lower, in the range $1 - 1.5$ for particles with harmonic (spring-like) interactions \cite{Olsson2007,Otsuki2009,Tighe2010,Vagberg2014}, which is believed to be a good model for emulsions \cite{Durian1995}. Second, while there is broad agreement that exponents for both static and flow properties are identical in two, three, and four dimensions \cite{OHern2003,Otsuki2009,Otsuki2009a,Katgert2013}, there is generally a dependence on particle interactions.} For example, when elastic quantities such as the confining pressure, shear modulus and yield stress are expressed as power laws in $\Delta \phi$, they scale with different exponents when the particles have harmonic or Hertzian interactions \cite{OHern2003,Heussinger2009}. 
This elastic force law dependence can be rationalized by recasting the scaling relations in dimensionless form, with units of stress constructed from contact-scale properties such as the stiffness $k$ between particles, generally depends on the proximity to jamming. {\bpt This approach differs from dimensionless quantities defined in terms of  particle-scale properties like the Laplace pressure.} How scaling exponents depend on the elastic force law is better understood than their dependence, if any, on the viscous force law. While different drag laws dramatically alter the form of correlation functions \cite{Tighe2010,Katgert2013,Baumgarten2017}, the Herschel-Bulkley exponent $\beta$ appears to be identical for the two most commonly used drag laws \cite{Vagberg2014}. 


{\bpt While experimental studies of jamming and rheology have focused on steady flow, simulations have also been used to probe oscillatory shear.} Recently, Dagois-Bohy et al. \cite{Dagois-Bohy2017} reported the softening and yielding of soft glassy materials, based on  athermal soft sphere simulations of both small and large amplitude oscillatory tests. They showed, amongst other things, scaling of the linear elastic and loss moduli of a two-dimensional system with respect to the quiescent confining pressure $P$ of the system. The results were in good agreement with theoretical models of small amplitude oscillatory shear \cite{Tighe2011,Baumgarten2017}. Data collapse into two master curves, one for the elastic modulus and one for the loss modulus, was found when plotting $G^*/P^\alpha$ versus $\omega / P^\beta$ with $\alpha \approx 0.45$ and $\beta \approx 0.8$, close to the mean field exponents $\alpha = 1/2$ and $\beta = 1$ \cite{Dagois-Bohy2017}. {\bpt These results also highlight the tendency to express numerical scaling relations in terms of the pressure, rather than excess volume fraction. This is advantageous because (i) the value of the confining pressure at jamming is strictly zero, and therefore known with arbitrary precision, and (ii) the pressure is easily accessible in simulations.}

The problem that we observe from the results summarized above, is the difference in scaling exponents between experiments and simulations. Experiments give scaling exponents that are independent of the details of the particle interactions \cite{Nordstrom2010,Paredes2013,Basu2014,Dinkgreve2015}. However, theoreticians observe interaction-dependent scaling of flow curves, i.e. different scaling exponents are found for harmonic and Hertzian particle interactions \cite{Olsson2007,Otsuki2009,Otsuki2009a,Tighe2010,Katgert2013,Vagberg2014,Baumgarten2017,Dagois-Bohy2017}. In this paper we try to answer the question how these differences in scaling exponents between experiments and simulations arise. Therefore, we look at three differences between experiments and simulations. Whilst experimental flow curves are commonly rescaled with the distance to jamming, simulation data are often rescaled with the confining pressure. We show the relation between the volume fraction and confining pressure and rescale our experimental data with respect to the confining pressure and compare the new scaling exponents with simulation values. Second, we compare oscillatory measurements, used in simulations, with rotational measurements, used in experiments and observe that above yielding, oscillatory and rotational flow curves overlap. Third, we have a closer look on the system's conditions. At this point we clearly observe a difference in the working regimes between experiments and simulations, explaining the discrepancy between the scaling exponents.  

\section{Materials and methods}	

\subsection{Castor oil-in-water emulsions}
Mobile castor oil-in-water emulsions stabilized by sodium dodecyl sulphate (SDS) were prepared. For the continuous phase, 1 wt\% of SDS (from Sigma-Aldrich) was dissolved in demineralized water. Rhodamine B (from Sigma-Aldrich) was added as a dye. The castor oil (from Sigma-Aldrich) was added to the aqueous phase and stirred with a Silverson 15 m-a emulsifier at 8,000 rpm for 6 minutes. The SDS solution was mixed with castor oil in a 1:4 volume ratio to obtain a 80 \% castor oil-in-water emulsion. This 80 \% emulsion was diluted with the SDS solution to obtain emulsions with lower $\phi$. All samples were centrifuged for 10 minutes at 1000 rpm to remove any air bubbles. The oil droplets have a mean diameter of 3.4 $\upmu$m with a polydispersity of 20 \%, determined using confocal laser scanning microscopy. 

\subsection{Rheological measurements}
The rheological measurements were performed on an Anton Paar MCR 302 rheometer. A 50 mm-diameter cone-and-plate geometry was used with a $1^{\circ}$ cone and roughened surfaces to avoid wall slip \cite{Bertola2003}. All samples were pre-sheared at a shear rate of 100 s$^{-1}$ for 30 s, followed by a rest period of 30 s to create a controlled initial state in all samples \cite{Paredes2013}. The rotational tests were performed by carrying out a shear rate sweep from 1,000 s$^{-1}$ to $1 \cdot 10^{-3}$ s$^{-1}$. The oscillatory measurements were performed by carrying out an amplitude sweep from 0.1 \% to $1 \cdot 10^{4}$ \% at a constant frequency of 1 Hz. A flow curve can be obtained from an oscillatory measurement, using that the shear rate is linear related to the strain and frequency, $\dot{\gamma} = \gamma \omega$.

\section{Results and discussion}

\subsection{Volume fraction versus confining pressure}
\begin{figure*}
	\centering
	\includegraphics[width=\columnwidth]{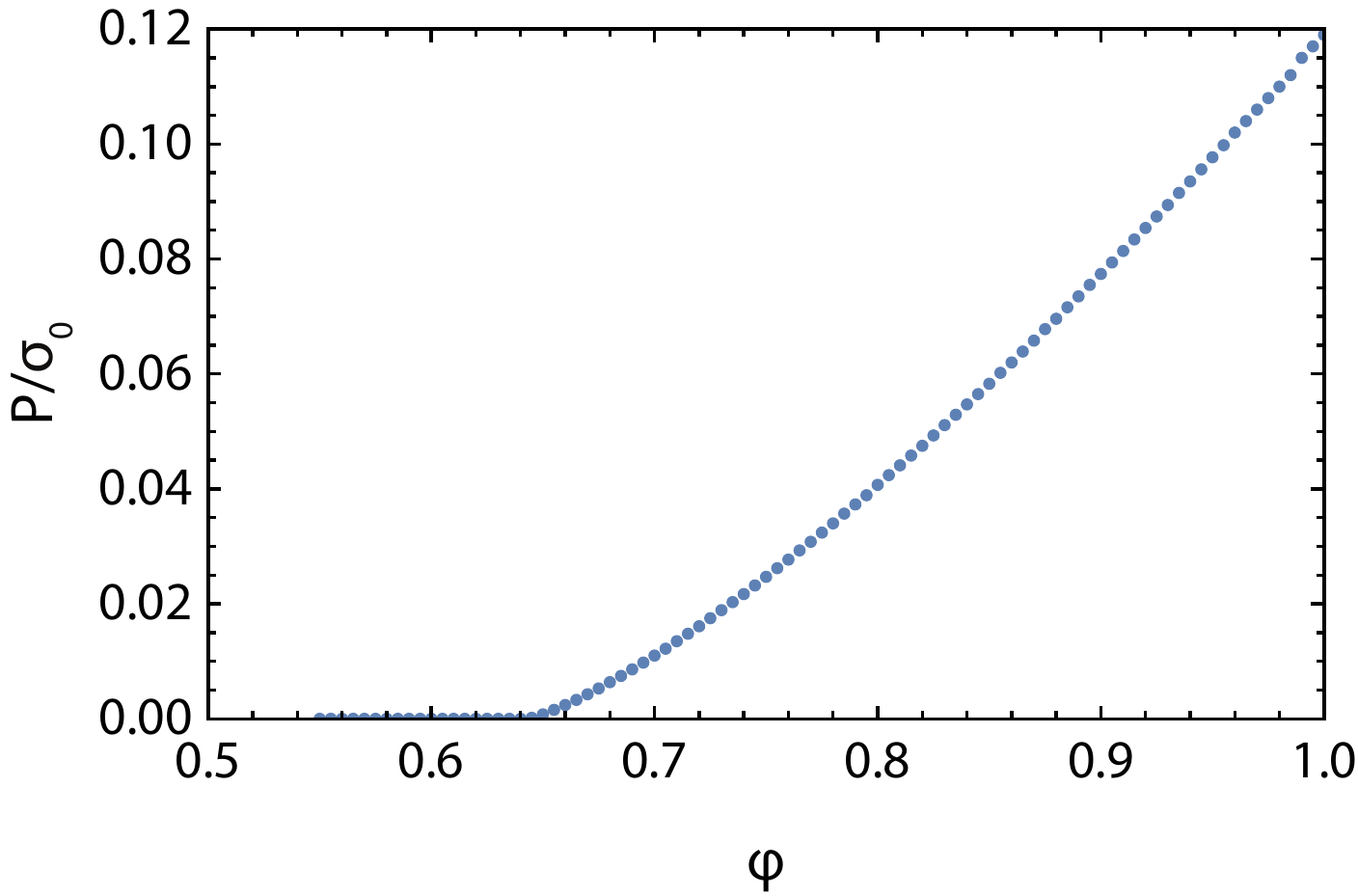}
	\caption{Numerical relation between the confining pressure and the volume fraction for a three-dimensional system. Below the critical volume fraction for jamming, $\phi_{c} = 0.64$, the pressure is zero.} 
	\label{volumefraction_pressure}
\end{figure*}

Whereas scaling of experimental flow curves is normally done with a power law in the distance to jamming, $\Delta \phi$, simulation data are scaled with a power law in the confining pressure. Relating the confining pressure with $\phi$ or $\Delta \phi$ would solve one of the discrepancies that might explain the differences in scaling exponents. A numerical relation between the quiescent (zero shear) confining pressure and the volume fraction for a three-dimensional system with a binary mixture of soft spheres is shown in Figure \ref{volumefraction_pressure}. Below the critical jamming point, $\phi_{c}=0.64$, the confining pressure is zero. While the initial growth of the pressure with $\Delta \phi$ is  linear \cite{OHern2003}, significant corrections to linear scaling are present over the experimentally accessible range of $\phi$, as can be seen from the slope of the graph. Therefore, we expect different effective exponents when scaling the flow curves with respect to the confining pressure in comparison with scaling with respect to $\Delta \phi$.

\begin{figure*}
	\centering
	\includegraphics[width=\linewidth]{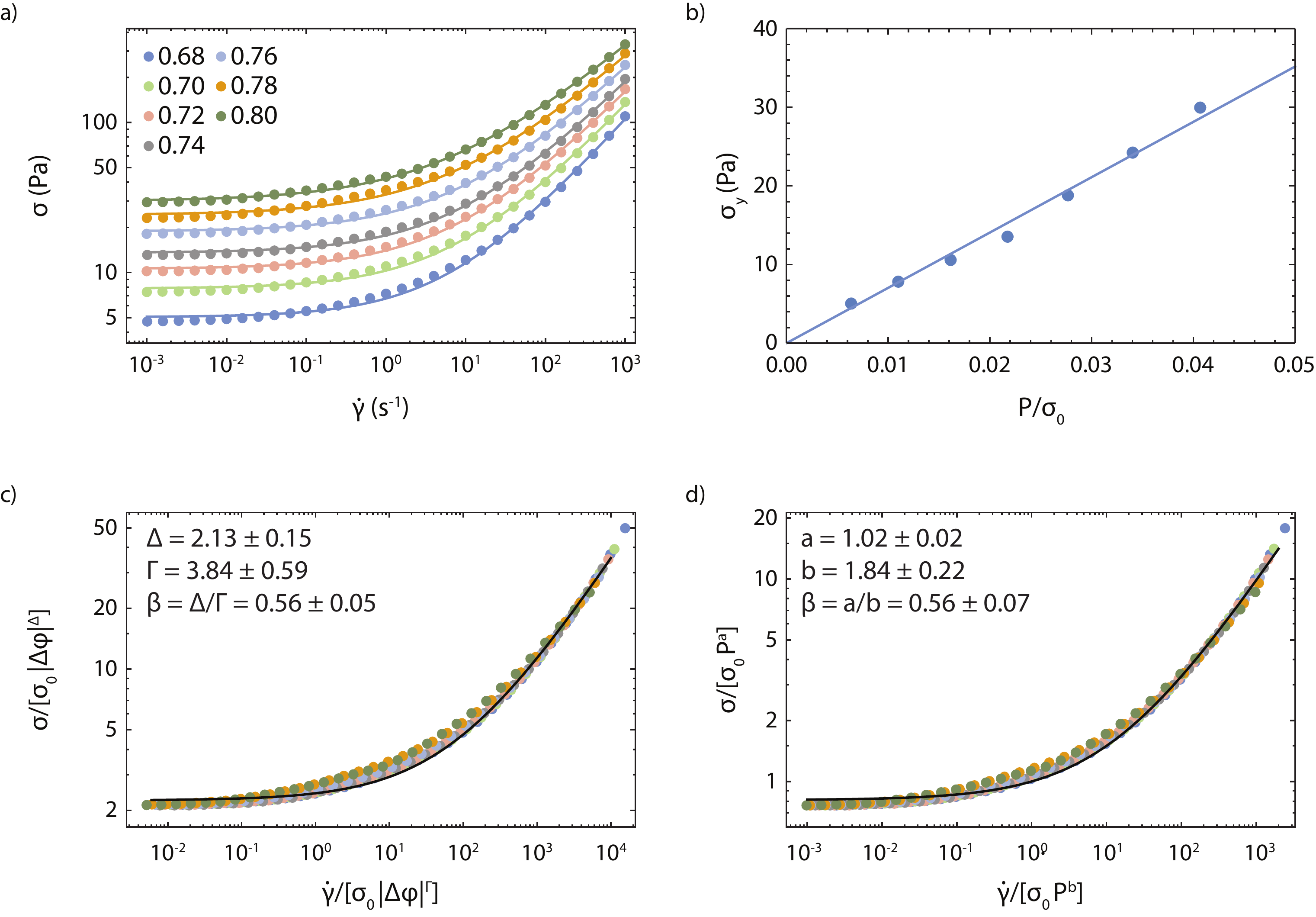}
	\caption{a) Flow curves of castor oil-in-water with 1wt\% SDS emulsions for volume fractions between 0.68 and 0.80 of the oil phase. The lines show Herschel-Bulkley fittings of the experimental data. b) Yield stress as a function of pressure with a linear fit through the origin. The yield stress is obtained from Herschel-Bulkley fits of the flow curves as shown in a). The volume fraction is related to the pressure using the numerical relation as shown in Figure \ref{volumefraction_pressure}. c) Master curve showing the collapse of the flow curves into one when plotted as $\sigma /|\Delta \phi|^{\Delta}$ vs $\dot{\gamma} /|\Delta \phi|^{\Gamma}$ with $\Delta = 2.13 \pm 0.15$ and $\Gamma = 3.84 \pm 0.59$. The black line corresponds to a Herschel-Bulkley fit with $\beta = a/b = 0.56$ and $K = 0.19$. d) Similar to c) but now plotted as $\sigma / P^{a}$ versus $\dot{\gamma} / P^{b}$ with $a = 1.02 \pm 0.02$ and $b = 1.84 \pm 0.22$. The black line corresponds to a Herschel-Bulkley fit with $\beta = a/b = 0.56$ and $K = 0.19$.}
	\label{flowcurves_rescaling}
\end{figure*}

Flow curves for castor oil-in-water emulsions with volume fractions between 68\% and 80\% are obtained from rotational measurements, see Figure \ref{flowcurves_rescaling}a. The data are fitted to the Herschel-Bulkley model, $\sigma = \sigma_{y} + K \dot{\gamma}^{\beta}$ {\bpt to determine the yield stress.} It can directly be seen from the curves in Figure \ref{flowcurves_rescaling}a that the yield stress increases with increasing volume fraction. The volume fractions were converted to confining pressure values using the numerical relation as shown in Figure \ref{volumefraction_pressure}. The values for the yield stress as obtained from the Herschel-Bulkley fits were plotted against the confining pressure, see Figure \ref{flowcurves_rescaling}b. The blue line shows a linear fit of the data points through the origin. The fit shows a linear relation between the yield stress and the confining pressure, with the yield stress vanishing at zero confining pressure. 

As discussed above, flow curves above the critical jamming volume fraction can collapse into one master branch by plotting $\sigma /\lvert \Delta \phi\rvert^{\Delta}$ vs $\dot{\gamma} /\lvert\Delta \phi\rvert^{\Gamma}$ \cite{Paredes2013,Dinkgreve2015}. Consistent with prior work, we find scaling collapse using $\Delta \approx 2.1$ and $\Gamma \approx 3.8$, see Figure \ref{flowcurves_rescaling}c. Critical scaling requires  $\beta = \Delta/\Gamma \approx 0.55$, and indeed a direct fit to the master curve gives fitting parameters $\beta = 0.56$ and $K = 0.19$. 


{\bpt We now attempt to rescale the  flow curves from Figure \ref{flowcurves_rescaling}a with the pressure, rather than $\Delta \phi$.}
The linear relation between the yield stress and the confining pressure, as shown in Figure \ref{flowcurves_rescaling}b, gives the first scaling parameter $a \approx 1$. We find that all data collapses into one master curve by plotting $\sigma /[\sigma_{0} P^{a}]$ versus $\dot{\gamma} /[\sigma_{0} P^{b}]$ with $a=1.02 \pm 0.02$ and $b=1.84 \pm 0.22$, see Figure \ref{flowcurves_rescaling}d. The master curve follows the Herschel-Bulkley model with $\beta = a/b = 0.56 \pm 0.07$. This shows that we can scale the flow curves with respect to the distance to jamming, but also with respect to the confining pressure. {\bpt Note that whereas $a \neq \Delta$ and $b \neq \Gamma$, their ratios $a/b$ and $\Delta/\Gamma$ (which set $\beta$) are the same. This is because, precisely at the jamming transition where both $P$ and $\Delta \phi$ are zero, the flow curve is that of a power law fluid $\sigma \propto \dot \gamma^\beta$. Hence $\beta$ cannot be sensitive to the choice to rescale with $P$ or $\Delta \phi$.}


\subsection{Rotational versus oscillatory measurements}
\begin{figure*}
\centering
\includegraphics[width=\linewidth]{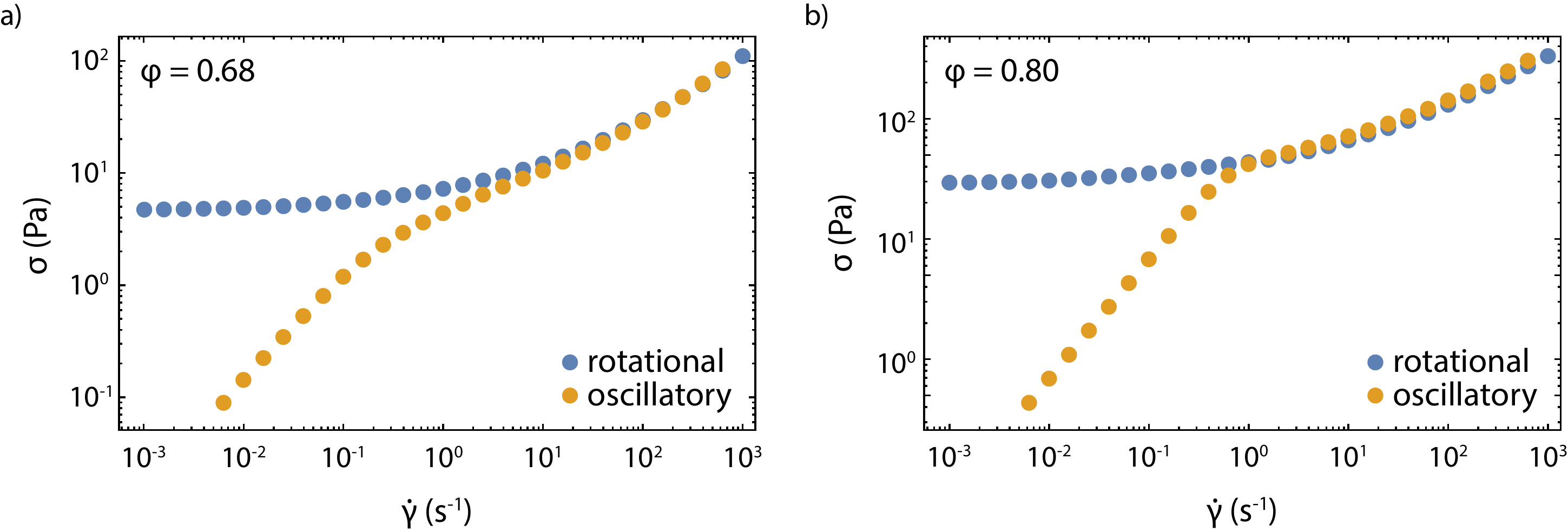}
\caption{Shear stress as a function of the shear rate for a castor oil-in-water with 1wt\% SDS emulsion with a volume fraction of a) 0.68 and b) 0.80 of the oil phase. The blue dots show a flow curve measured with a rotational test, whereas the orange dots show a flow curve measured with an oscillatory test.} 
\label{flowcurves_rotational_oscillatory}
\end{figure*}

Rotational measurements are the most common way in experimental research to obtain flow curves of emulsions. A shear rate is applied and the stress is measured. However, scaling of simulation data is often applied to the storage and loss moduli of a system \cite{Tighe2011,Dagois-Bohy2017}, information that can be obtained from oscillatory measurements. These measurements also give stress versus shear rate flow curves, which allows us to compare both methods. Flow curves were obtained for castor oil-in-water emulsions with volume fractions of the oil phase between 68\% and 80\% both from rotational and oscillatory measurements. The results for the emulsions with the lowest (68\%) and the highest (80\%) volume fractions are shown in Figure \ref{flowcurves_rotational_oscillatory}. The graphs show a clear overlap between the curves of the rotational and the oscillatory tests at high shear rates ($\dot{\gamma} \gtrsim 1 s^{-1}$) above yielding {\bpt (recall that we vary $\dot \gamma$ by sweeping strain amplitude at fixed frequency)}. The difference between the flow curves at lower shear rates reflects insufficient strain in the oscillatory measurements to reach a steady flow \cite{Dinkgreve2017}. The first part of the oscillatory measurements shows purely elastic behaviour of the sample. From the overlap between both flow curves at high shear rates, we can conclude that the same information can be obtained from rotational and oscillatory measurements. 

\subsection{Elastic versus viscoelastic material properties}
{\bpt We showed above that flow curves of oscillatory measurements overlap with flow curves of rotational measurements in the regime above yielding. Can we glean further information from small amplitude response? The scaling exponents for soft spheres in small amplitude oscillatory shear can be calculated theoretically if one assumes that the strain amplitude $\gamma$ is small enough to neglect all rearrangements between droplets \cite{Tighe2011}. In practice, the theoretically-predicted exponents remain valid even in the presence of a small amount of rearrangement events \cite{Boschan2016,Dagois-Bohy2017}. A detailed discussion of the experimental implications of softening and yielding was given by Boschan et al. \cite{Boschan2016}. 

Many complex fluids satisfy the empirical Cox-Merz rule $\sigma(\dot \gamma) = |G^*(\omega)|_{\omega = \dot \gamma}$, which relates the steady state flow curve $\sigma(\dot \gamma)$ to the complex shear modulus $G^*(\omega)$ in oscillatory measurements \cite{Barnes1989}. Combining the theory of small amplitude oscillatory shear in soft spheres \cite{Tighe2011,Baumgarten2017} with the Cox-Merz relation predicts $a = 1/2$ and $b = 1$. This is not in agreement with the values $a \approx 1$ and $b \approx 2$ determined above. However it is interesting to note that the Herschel-Bulkley exponent $\beta = a/b = 1/2$ is again robust. In small amplitude response the value of $\beta$ is directly related to an abundance of slow, strongly non-affine relaxational modes, which emerge near jamming \cite{Tighe2011}.  We speculate that yielding-induced rearrangements serve as an alternate source of non-affinity in steady flow.}

\begin{figure*}
	\centering
	\includegraphics[width=\linewidth]{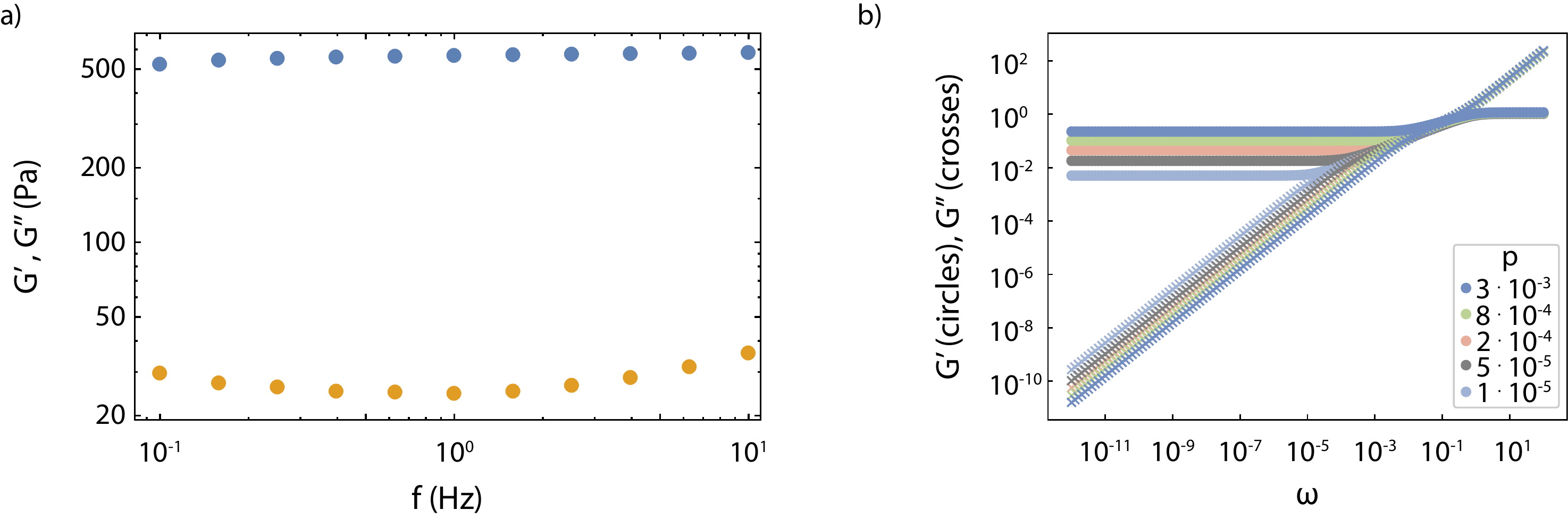}
	\caption{a) Experimental storage (blue circles) and loss (orange circles) moduli as a function of frequency for a 80 \% castor oil-in-water with 1wt\% SDS emulsion. At low frequencies, the storage and loss modulus are independent of frequency. b) Numerical storage (circles) and loss (crosses) moduli as a function of frequency $\omega$ for various confining pressures close to jamming. An increase in the loss modulus is observed with increasing frequency, whereas the storage modulus remains constant in a large range of frequencies.} 
	\label{oscillation-measurements}
\end{figure*}

A direct comparison of oscillatory measurements in experiments and simulations reveals important differences. As can be seen in Figure \ref{oscillation-measurements}a, $G'$ and $G''$ for a 80\% castor oil-in-water emulsion are independent of the frequency in the low frequency regime. {\bpt (Except at the critical point, mathematical properties of the Fourier transform require $G''$ to vanish linearly at small frequencies. Presumably this occurs outside the experimental window.)} Numerical data from soft sphere systems at various confining pressures show no such plateau in $G''$, see Figure \ref{oscillation-measurements}b. Whilst the storage modulus remains constant with increasing frequency, the loss modulus increases linearly. {\bpt To the best of our knowledge, the origin of this discrepancy between experimental and numerical loss moduli remains unexplained.}

\section{Conclusion}
In this article we asked where the differences in scaling exponents between experimental and simulation flow curves stem from. {\bpt A key difference regards how closely experiments and simulations are able to approach the jamming point, with simulations  able to probe much closer to jamming. As a result the two techniques typically probe different windows of response. Experimental results can display significant corrections to scaling, as demonstrated by our finding that scaling with pressure and excess volume fraction give different exponents over experimentally accessible parameter ranges. Intriguingly, we also showed that oscillatory and rotational measurements give the same flow curves above the yielding point, pointing to connections between steady flow and large amplitude oscillatory shear. Finally we found that the Herschel-Bulkley exponent $\beta$ is remarkably robust, with a value of approximately 0.5 for different experimental systems and for different choices of scaling parameter, and even comparing favorably between steady flow and small amplitude oscillatory shear. On the other hand, unexplained phenomena remain, including the value of the scaling exponent $\Delta$ within the experimental window, and the origin of the plateau in $G''$. }


\bibliographystyle{promotie-riande}
\bibliography{C:/Users/Riande/Documents/Writing/library}

\end{multicols}
	
\end{document}